\documentstyle[epsfig]{europhys}

\def\etal{{\hbox{{\tenit\ et al.\/}\tenrm :\ }}}

\def\stars{\bigskip\centerline{***}\medskip}

\newif\ifboo \boofalse

\begin{document}
%
%
%
\euro{}{}{}{}
\Date{}
\shorttitle{Mahn-Soo Choi\etal Anderson-type model}
%
%
%
\title{Anderson-type model for a molecule adsorbed on a metal surface}
\author{Mahn-Soo Choi and C. Bruder}
\institute{
   \inst{1} Departement Physik und Astronomie, Universit\"at Basel,
Klingelbergstrasse 82, CH-4056 Basel, Switzerland}
%
%
\rec{}{}
%
%
%
\pacs{
\Pacs{73}{40Gk}{Tunneling}
\Pacs{85}{65$+$h}{Molecular electronic devices}
\Pacs{61}{16Ch}{Scanning probe microscopy}
}
%
\maketitle
%
%
%
\begin{abstract}
We investigate a modified Anderson model to study the local density of
states (LDOS) of a molecular wire adsorbed on a metal. Using a
self-consistent mean-field type approach we find an exponential decay
of the LDOS along the molecule. A repulsive on-site interaction on the
molecule suppresses the tunneling and decreases the characteristic
decay length.
\end{abstract}
%
\newcommand\up{\uparrow}
\newcommand\down{\downarrow}
\newcommand\beq{\begin{equation}}
\newcommand\eeq{\end{equation}}
\newcommand\beqa{\begin{eqnarray}}
\newcommand\eeqa{\end{eqnarray}}
\newcommand\beqaz{\begin{eqnarray*}}
\newcommand\eeqaz{\end{eqnarray*}}
\newcommand\lavg{\left\langle\:}
\newcommand\ravg{\:\right\rangle}
\newcommand\bfk{{\bf k}}
\newcommand\varG{{\mathcal{G}}}
\newcommand\hatd{{\hat{d}}}
\newcommand\hatG{{\widehat{G}}}
\newcommand\im{{\rm Im}}
\newcommand\re{{\rm Re}}
\newcommand\egg{{}}

\def\etal{{\hbox{{\it\ et al.\/}\rm :\ }}}


\section{Introduction}
In the last decade, a lot of progress has been made in the theoretical
and experimental investigation of nanostructures like quantum dots and
quantum wires \cite{curacao}. These structures are produced by
lithography. Using this very flexible technology, a large variety of
different structures with different geometries has been made. Simple
dots, multiple dots, dot arrays and the corresponding leads have been
realized both using metals and semiconductors. A disadvantage of
lithographic methods is the presence of fluctuations in shape and
size of these structures; they are unavoidable because of errors in
writing the mask and because of the stochastic nature of the etching
process.

The idea to use molecules or supramolecular structures as quantum
wires and quantum dots has been around for quite a while, see
Ref.~\cite{aviram74} for an early suggestion of a molecular rectifier
and Refs.~\cite{petty95,aviram98} for recent reviews. Powerful
chemical synthesis methods are available that allow the production of
atomic and molecular clusters with linear dimensions of up to 5 nm,
i.e., they are approaching the lowest linear dimensions of
nanostructures produced by lithography. It is a fascinating idea that
such clusters which would have custom-made electronic properties could
be included in nanostructures and replace, say, a quantum
dot. Different copies of these systems would be identical in structure
and properties (e.g., there is no difference in the composition of
different C$_{60}$ molecules). First steps in the direction of this
`bottom-up approach' are under way in a number of labs around the
world. Scanning probe techniques have been used to arrange atoms on a
substrate in arbitrary ways \cite{eigler,avouris}. A particularly
impressive example is the `quantum corral' \cite{crommie}, a
circularly shaped structure of Fe atoms on a Cu(111) surface that may
be considered a precursor of a device. Carbon nanotubes have been
contacted and proven to exhibit Coulomb-blockade behavior
\cite{dekker1,bachtold}. A new kind of transistor whose central
element is a C$_{60}$ molecule has been proposed
\cite{gimzewski98}. The theory of conduction through molecular wires
has been studied by a number of groups (see, e.g.,
Refs. \cite{ratner94,joachim96,magoga9798,datta,kirczenow}). Also,
the thermal conduction through molecular wires has been investigated
\cite{ciraci}.  Recently, the coupling of a molecule to a metal has
been studied both experimentally and theoretically
\cite{kergueris,dattapaper}.

A central problem facing the use of molecules as electronic
devices is the question of how to contact them in a reproducible way.
An interesting answer to this question was given in a recent
experiment \cite{gimzewski99} in which a custom-designed ``lander'' molecule
\cite{gourdon98} was adsorbed on a $(111)$ copper surface. The
molecules can be thought of as consisting of an aromatic platform and
four spacer ``legs''. The spacer legs keep the main board at such a
distance from the metal surface that the aromatic part is
electronically decoupled from the surface.

The molecules self-assembled perpendicularly to a step on the surface,
thereby forming a contact between the metal and the aromatic board
(the length of the legs was chosen to be comparable to the step height). A
scanning-tunneling microscope (STM) was used to scan along the surface
and the molecule, and it was found that the tunneling conductance
decayed in an exponential way along the molecule. 

In this paper, we would like to study a (relatively) simple model for
the molecule at the metal step. In contrast to the quantum-chemical
calculations provided in \cite{gimzewski99} that take into account the
precise structure of the lander molecule, we propose to use a modified
Anderson model to describe this situation. The `impurity' of the new
model is spatially extended, i.e., contains internal degrees of
freedom not present in the standard Anderson model. The aim will be to
calculate the local density of states of an extended structure (the
molecule) using a Green's function method. The molecule will be
approximated by a (finite) one- or two-dimensional tight-binding
lattice with a Hubbard-like interaction. Our goal is to gain a {\it
qualitative understanding} of the physical problem and its
phenomenology (as opposed to quantum-chemical calculations for {\it
specific} molecules) to be able to propose and analyze more
complicated (e.g., multiply-connected) structures in the future.
\begin{figure}\centering
\hbox{\epsfig{file=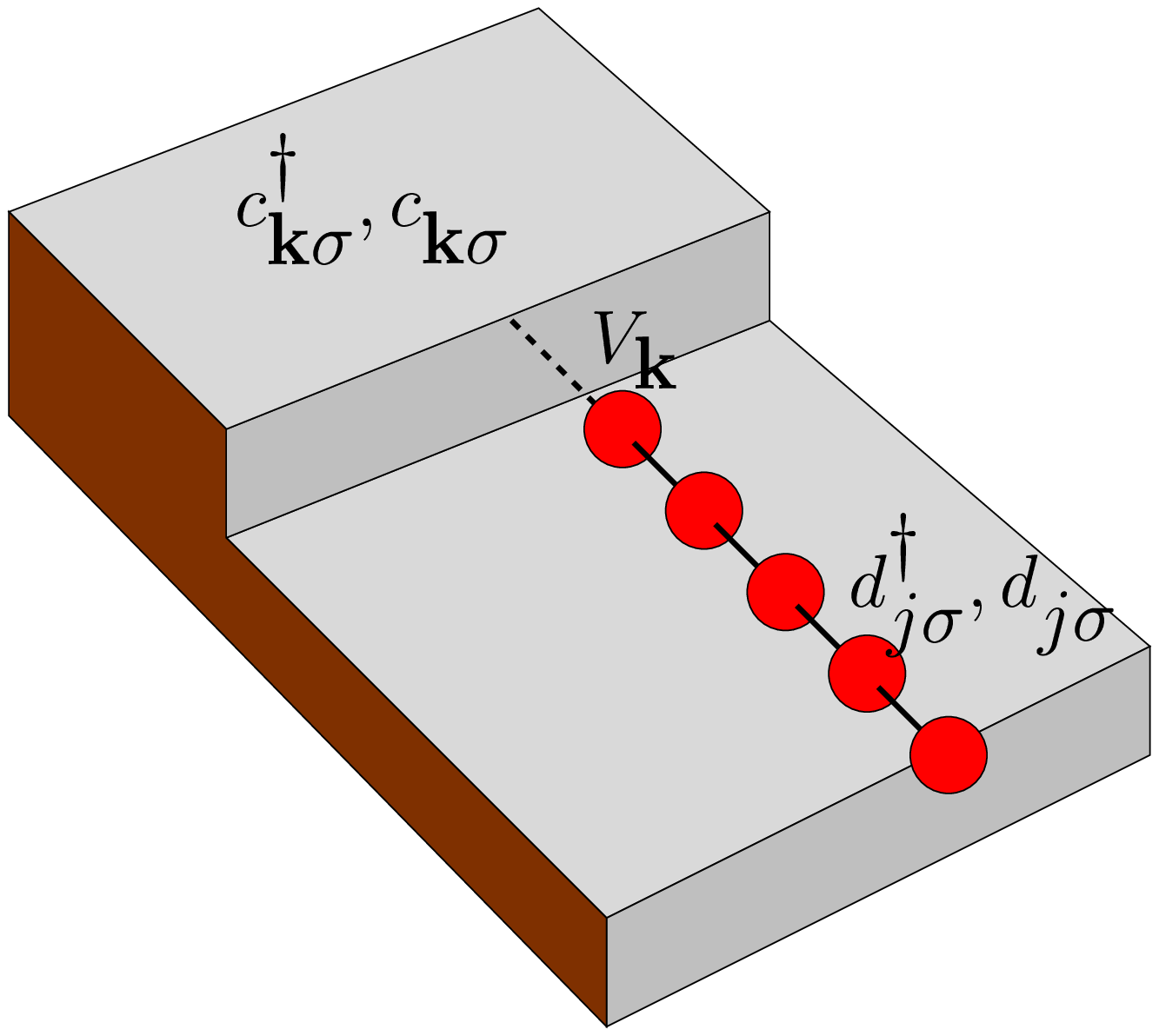,clip=,height=4.5cm}
\epsfig{file=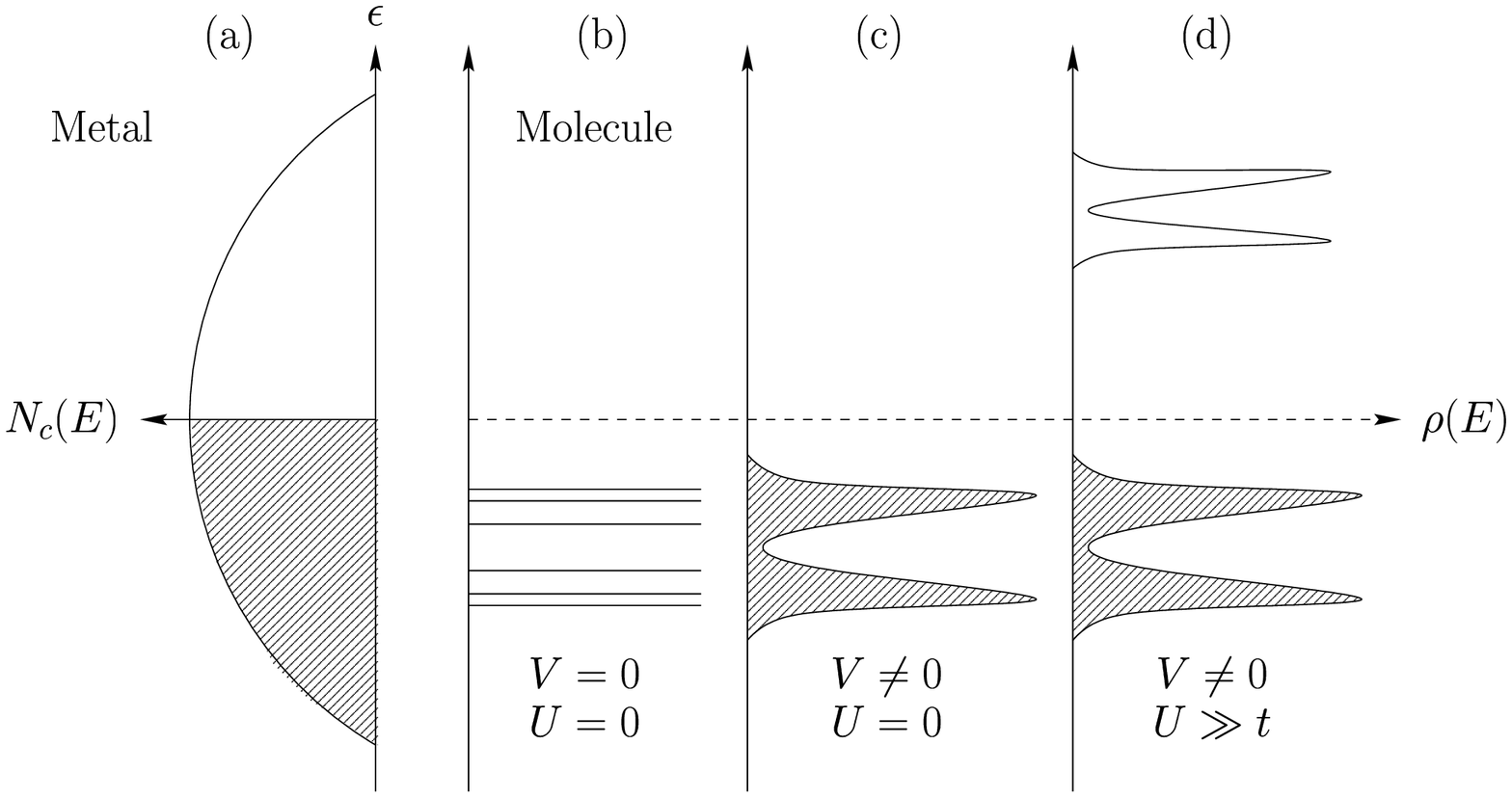,clip=,height=4.5cm}}
\caption{Left part: schematic view of the system. We describe the molecule 
by a tight-binding model, i.e., a discrete lattice of atomic sites.
The coupling of the molecule to a metal consisting of
non-interacting electrons in a conduction band is assumed to be small
and characterized by the tunneling amplitude $V_\bfk$.
Right part: sketch of the energy spectrum of the model. (a)
unperturbed density of states $N_c(E)$ of the metal. (b) - (d)
density of states $\rho(E)$ at a typical molecular site. (b) Isolated
molecule ($V_\bfk=0$) without electron-electron interaction
($U=0$). (c) Molecule without interaction coupled to
the metal. The coupling broadens the discrete molecular levels. (d)
Molecule in the presence of strong electron-electron interaction
($U\gg{t}$) coupled to the metal. In the half-filled case (one
electron per site), the interaction will lead to the formation of an
upper Hubbard band.}
\label{moms:fig1}
\end{figure}

\section{Model}
The system we have in mind is shown in Fig.~\ref{moms:fig1}.
Our model is defined by a Hamiltonian that consists of three parts, 
$
H=H_c+H_d+H_T
$.
The metal to which the molecular
wire is attached is described by non-interacting electrons in a conduction band
\beq\label{moms:1}
H_c
= \sum_{\bfk\sigma}\epsilon_\bfk c_{\bfk\sigma}^\dag
c_{\bfk\sigma}^{}\; ,
\eeq
where $\bfk$ and $\sigma=\up,\down$ denote electron wave vectors and spins,
respectively. The single-electron energy $\epsilon_\bfk$ is measured with
respect to the Fermi energy ($\epsilon_F=0$).
The molecular wire is assumed to be a one-dimensional
lattice of $L$ atomic sites and is described within the Hubbard model
\beq\label{moms:2}
H_d
= \sum_{i,j=1}^Lt_{ij}d_{i\sigma}^\dag d_{j\sigma}^\egg
 + U\sum_{j=1}^Ln_{j\up}n_{j\down}\; .
\eeq
The parameters that enter here are the on-site electron-electron
interaction $U$ and the matrix elements
$t_{ij}=\epsilon_d\delta_{ij}-t\,\delta_{i,j\pm1}$, where $\epsilon_d$
is the on-site energy of the single-particle level at each site and
$t$ is the hopping element between nearest neighbors.
The coupling of the wire to the metal is assumed to be small so that we can
describe it by the tunneling amplitude $V_\bfk$ of the electron in the
state $(\bfk,\sigma)$ to the first site $j=1$:
\beq\label{moms:3}
H_T
= -\sum_{\bfk\sigma}\left(V_\bfk c_{\bfk\sigma}^\dag d_{1\sigma}^\egg
 + h.c.\right)\; .
\eeq
The model defined by $H$
 is a generalization of 
the Anderson impurity model: for $L=1$, the molecule reduces to a
localized level, and (\ref{moms:3}) is the hybridization term of the
Anderson impurity model. 

The qualitative nature of the spectrum of the system is shown in the
right part of Fig.~\ref{moms:fig1}. We characterize the metal by a
model density of states $N_c(E)$ to be discussed below and shown
schematically in (a). If the molecule is not
coupled to the metal, $V=0$, its spectrum consists of discrete levels
as shown in (b). Coupling the molecule to the metal broadens the
spectral lines as shown in (c). If we assume a filling of one electron
per site, the presence of strong on-site interactions leads to the
formation of a Mott-Hubbard gap as shown in (d).

We will use the imaginary-time path-integral formalism at finite
temperature $k_BT\equiv1/\beta$ (see, e.g., Ref.~\cite{Negele88}).
In this formalism, the Euclidean action is given by
\beq\label{moms:6}
S^E
= \int_0^\beta{d\tau}\;\left(
 \sum_{\bfk\sigma}c_{\bfk\sigma}^\dag\partial_\tau\,c_{\bfk\sigma}^\egg
 + \sum_{j\sigma}d_{j\sigma}^\dag\partial_\tau\,d_{j\sigma}^\egg
 + H
 \right) \; ,
\eeq
and we want to calculate the thermal Green's function
\beq\label{moms:a1}
\varG(j\sigma,k\sigma';\tau)
= -\lavg T_\tau d_{j\sigma}^\dag(0) d_{k\sigma'}(\tau) \ravg
\eeq
and its Fourier transform defined by 
$\varG(j\sigma,k\sigma';i\omega_n)
 = \int_0^\beta{d\tau}\;e^{i\omega_n\tau}\varG(j\sigma,k\sigma';\tau)
$,
where $\omega_n\equiv(2n+1)\pi/\beta$ with $n$ integer are the Matsubara
frequencies.

First, we consider the non-interacting case, $U=0$. In this case, the
Euclidean action (\ref{moms:6}) is quadratic both in $c_{\bfk\sigma}^\egg
(c_{\bfk\sigma}^\dag)$ and $d_{j\sigma}^\egg (d_{j\sigma}^\dag)$, and it
follows that
$\varG(j\sigma,k\sigma';i\omega_n)=
\delta_{\sigma\sigma'}\varG(j,k;i\omega_n)$,
\beq\label{moms:7}
\varG^{-1}(j,k;i\omega_n)
= \varG_0^{-1}(j,k;i\omega_n)-\Sigma_c(j,k;i\omega_n)\; ,
\eeq
or more explicitly
\beq\label{moms:8}
\varG(i,j)
= \varG_0(i,j)
+
\frac{\varG_0(i,0)\Sigma_c(0,0)\varG_0(0,j)}{1-\varG_0(0,0)\Sigma_c(0,0)}\; .
\eeq
The unperturbed (non-interacting, isolated) Green's function is given by
\beq\label{moms:9}
\varG_0^{-1}(j,k;i\omega_n)
= i\omega_n - t_{jk}\; .
\eeq
In Eqs.~(\ref{moms:7}) and (\ref{moms:8}), the coupling to the metal
manifests its effect on the wire through the self-energy
\beq\label{moms:a2}
\Sigma_c(j,k;i\omega_n)
= \delta_{j1}\delta_{k1}\,
 \int_{-\infty}^\infty\frac{d\epsilon}{2\pi}\;
 \frac{\Gamma(\epsilon)}{i\omega_n-\epsilon}\; .
\eeq
The effect of the metal on the molecular wire is given 
by the function
\beq\label{moms:4}
\Gamma(E)
\equiv -2\sum_\bfk |V_\bfk|^2 \im G_c^R(\bfk,E)\; ,
\eeq
where $G_c^R(\bfk,E)$ is the retarded Green's function for the
(unperturbed) metal. In the simplest case in which the
energy-dependence of the tunneling amplitude $V_\bfk$ is not too
large, $V_\bfk\approx V$, (\ref{moms:4}) can be further reduced to an
expression directly proportional to the density of states in the metal
$N_c(E)$: $\Gamma(E) = 2\pi V^2N_c(E)$.

The local density of states (LDOS) at the $j$th site is then given by the
analytic continuation of the thermal Green's function.
\beq\label{moms:10}
\rho(j,E)
= -\frac{2}{\pi}\im\varG(j,j;i\omega_n\to E+i0^+)\; .
\eeq
The factor 2 accounts for the contributions from the two spin components.

We now turn to the opposite limit in which the electron-electron
interaction on the molecule is very strong; $U\gg{t}$
\cite{dimerization}. In this case, exact solutions are only available
in special cases; the isolated molecule ($V_\bfk=0$), i.e, the usual
Hubbard model \cite{Liebxx68} and the single-site molecule ($L=1$),
i.e., the usual Anderson impurity model \cite{Wiegma81}. There are
many approximation methods for the Hubbard model or the Anderson model
(see Ref.~\cite{George96} for a recent review), and we will adopt a
self-consistent mean-field approximation \cite{Negele88}.

We start with the strong-$U$ limit of $H_d$ in (\ref{moms:2}), i.e., the 
$t{-}J$ Hamiltonian \cite{Auerba94}
\beq\label{moms:11}
H_d
\simeq \sum_{ij}t_{ij}d_{i\sigma}^\dag d_{j\sigma}^\egg
- \frac{1}{U}\sum_{ijk} 
 t_{ij}(d_{i\up}^\dag d_{j\down}^\dag - d_{i\down}^\dag d_{j\up}^\dag)
 t_{jk}(d_{j\down}^\egg d_{k\up}^\egg - d_{j\up}^\egg d_{k\down}^\egg)
 \nonumber
\eeq
within the reduced Hilbert space without doubly occupied sites.
The second term in (\ref{moms:11}) can be made quadratic in
$d_{j\sigma}^\egg (d_{j\sigma}^\dag)$ by means of the
Hubbard-Stratonovich transformation introducing the auxiliary field
$\Delta_j$ to get the molecular part (corresponding to $H_d$) in the
Euclidean action (\ref{moms:6})
\beq\label{moms:12}
S_d^E
= S_\Delta^E
+ \int_0^\beta{d\tau}\sum_{ij}
 \hat{d}_i^\dag\left[
 \delta_{ij}\partial_\tau
 + t_{ij}\hat\tau_3 - \widehat\Delta_{ij}
 \right]\hat{d}_j^\egg\; ,
\eeq
where $\tau_3$ is the Pauli matrix and 
$
S_\Delta^E
= \int_0^\beta{d\tau}\;U\sum_j|\Delta_j|^2
$.
%
In Eq.~(\ref{moms:12}), we have introduced a two-component spinor
representation \beq \hat{d}_j \equiv \left[\begin{array}{c}
d_{j\up}^\egg\\d_{j\down}^\dag \end{array}\right] ,\quad
\widehat\Delta_{ij} \equiv t_{ij}\left[\begin{array}{cc} 0 &
\Delta_i^\egg+\Delta_j^\egg \\ \Delta_i^*+\Delta_j^* & 0
\end{array}\right]\; . \eeq In this representation, the Green's
function in Eq.~(\ref{moms:a1}) also has a matrix form
$\widehat\varG(i,j;\tau) = -\lavg
T_\tau\hat{d}_i^\egg(0)\hat{d}_j^\dag(\tau)\ravg $.
After integrating out the fields $c_{\bfk\sigma}^\egg$ and
$c_{\bfk\sigma}^\dag$, the Euclidean action can be written as
\beq\label{moms:14}
S^E
= S_\Delta^E
- \int\!\!\!\int_0^\beta{d\tau d\tau'}\;
 \hatd_i^\dag(\tau)\widehat\varG^{-1}(i,j;\tau-\tau')
 \hatd_j^\egg(\tau')
\eeq
where the Green's function $\varG$ is given by the Dyson equation
\beq\label{moms:15}
\widehat\varG^{-1}(j,k;i\omega_n)
= \widehat\varG_0^{-1}(j,k;i\omega_n)
 - \widehat\Sigma_c(j,k;i\omega_n)
 - \widehat\Sigma(j,k;i\omega_n)\; ,
\eeq
with the new self-energy term
$\widehat\Sigma(i,j;\tau)=-\delta(\tau)\widehat\Delta_{ij}(\tau)$.
Here $\varG_0$ in (\ref{moms:9}) and $\Sigma_c$ in Eq.~(\ref{moms:a2}) have
been extended to matrix form in a trivial way:
$\widehat\varG_0=\varG_0\hat\tau_0$ and
$\widehat\Sigma_c=\Sigma_c\hat\tau_0$, where $\tau_0$ is the identity
matrix.

So far no approximation has been made and the formal expression for
the Euclidean action in Eq.~(\ref{moms:14}) is exact. The interaction
effect is correctly incorporated through the self-energy term
$\widehat\Sigma$ (whereas the effect of the metal is again manifested
by $\widehat\Sigma_c$) as long as the integration over the field
$\Delta_j$ is performed properly. At this point, we make our main
approximation and neglect the fluctuations in the field $\Delta_j$. We
first assume a particular realization of the field $\Delta_j$ in the
Dyson equation (\ref{moms:15}) and then determine $\Delta_j$ within
the stationary-phase approximation. Iterating this procedure allows
us to determine $\Delta_j$ and the Green's function self-consistently.
Moreover, the self-consistency equation
\beq\label{moms:16}
\Delta_i
= \frac{1}{U}\sum_j t_{ij}\lavg d_{i\down}d_{j\up}-d_{i\up}d_{j\down} \ravg
\eeq
corresponding to this procedure, gives us a physical interpretation of
the field $\Delta_j$: $\Delta_j$ measures the spin-singlet correlation
between the $j$th site and its neighboring sites. It is clear that in
the limit $U\to\infty$, $\Delta_j\approx2t/U$ unless $j$ is too close
to the boundaries. Finally, the LDOS can be obtained from the
diagonal components of the retarded Green's function
$\hatG^R(j,k;E)=\widehat\varG(j,k;i\omega_n\to E+i0^+)$:
\beq
\rho(j,E)
= -\frac{1}{\pi}\im\left[\hatG_{11}^R(j,j;E)+\hatG_{22}^R(j,j;-E)\right]\; .
\eeq
\section{Results}
We will now report specific results on the LDOS of the
molecular wire and discuss their physical implications. For the
density of states (DOS) in the metal, we adopt the model functional
form 
$
N_c(E)
= N_c(0)(1-E^2/W_c^2)
$.
The parameters $N_c(0)$ and $W_c$ specify the DOS at the Fermi energy
and the width of the conduction band, respectively.
The self-energy contribution (\ref{moms:a2}) from the coupling to the
metal is then given by
\beq\label{moms:18}
\Sigma_c(j,k;E)
= \delta_{j1}\delta_{k1}\,V^2
 \left[ \alpha_c(E)-i\pi N_c(E) \right]
\eeq
with $\alpha_c(E)$ defined by
\beq\label{moms:19}
\alpha_c(E)
= N_c(0)\left[
 \frac{2E}{W_c}
 + \left(1-\frac{E^2}{W_c^2}\right)
  \ln\left|\frac{E/W_c+1}{E/W_c-1}\right|
 \right] \;.
\eeq
All results shown in this work will be generated using the values
$\epsilon_d=-4t$, $V=0.1t$, $W_c=40t$, and $N_c(0)=10/t$. The other
parameters will be specified as needed. We also remark that we only
investigate the lower band, i.e., the band below the Mott-Hubbard gap,
in the interacting case.

\begin{figure}\centering
\epsfig{file=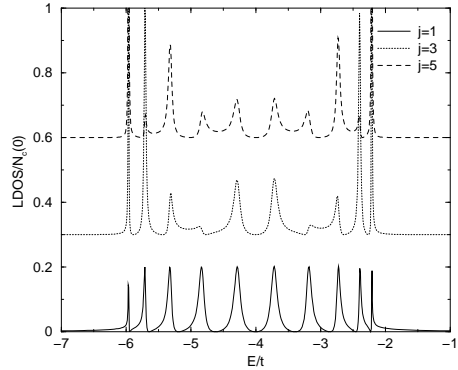,clip=,width=6cm}
\caption{Local density of states at site $j$ for $L=10$ and $U=10t$.
Solid line: $j=1$, dotted line: $j=3$ (shifted upwards by $0.3$),
dashed line: $j=5$ (shifted upwards by $0.6$).}
\label{moms:fig3}
\end{figure}

In Fig.~\ref{moms:fig3} we plot typical LDOS curves at three different
sites ($j=1$, $3$, and $5$) on a molecule with size $L=10$ and on-site
interaction $U=10t$. The structures get sharper as $U$ increases
\cite{Brinkm70} because tunneling is blocked, see the discussion
below. At a given site, the LDOS decays rapidly outside the band.

\begin{figure}\centering
\epsfig{file=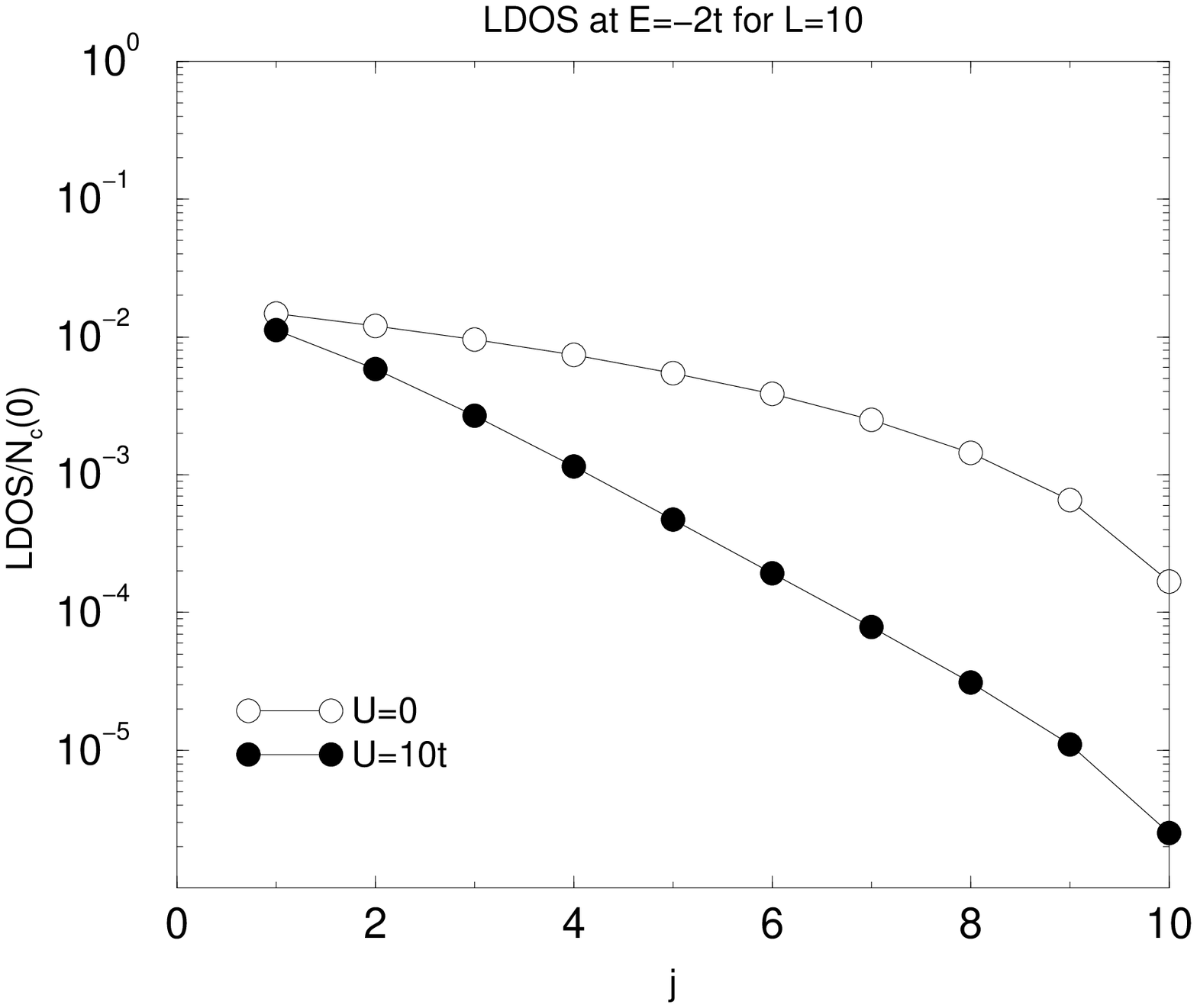,clip=,height=5cm}
\epsfig{file=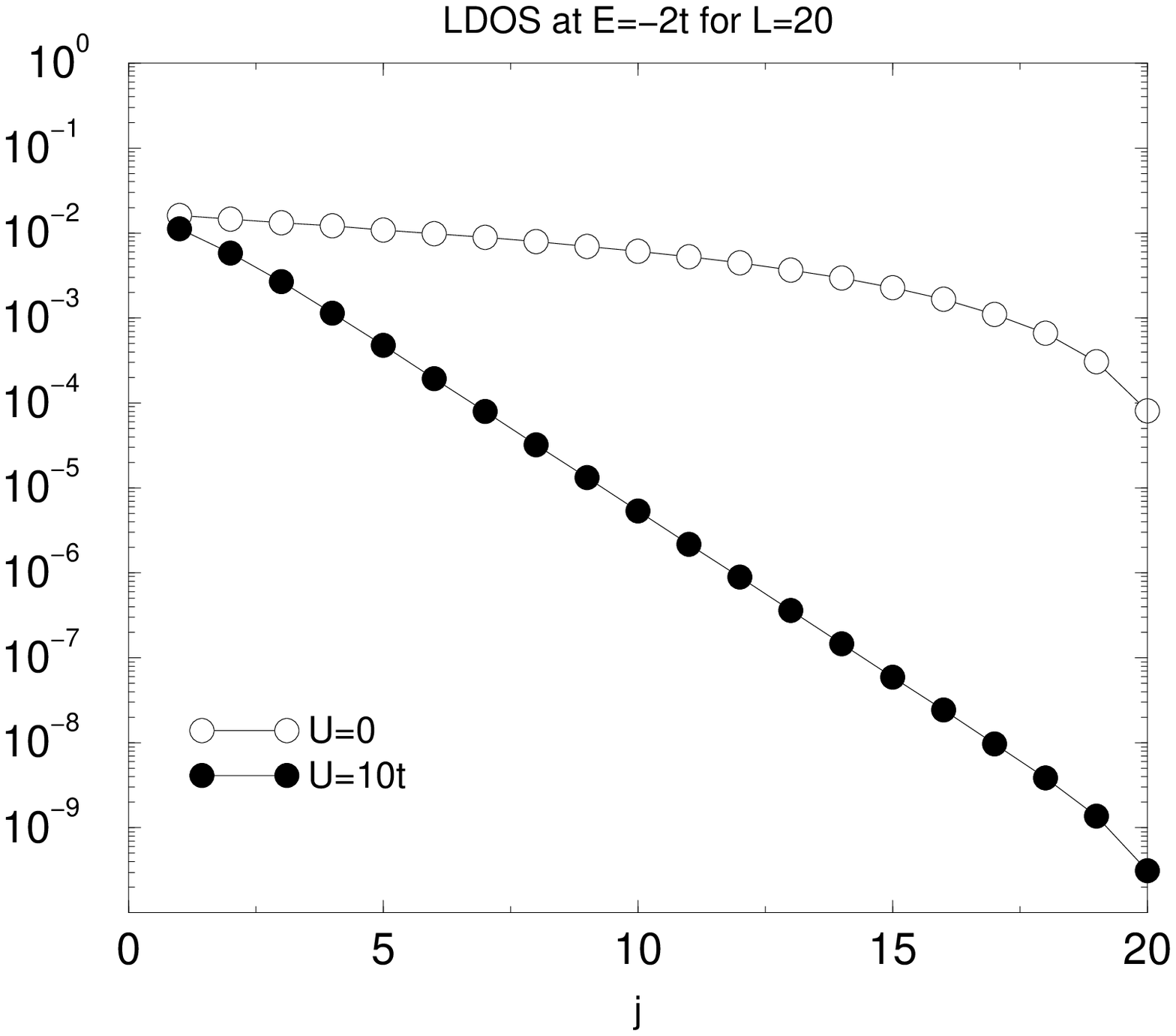,clip=,height=5cm}
\caption{Spatial dependence of the local density of states along the
molecule. Empty symbols: non-interacting case. Full symbols:
interacting case, $U=10t$. Interaction effects on the molecule depress
the influence of the metals on the density of states on sites that are
far away.}
\label{moms:fig4}
\end{figure}

Figure \ref{moms:fig4} shows the spatial dependence of the LDOS along
molecules with sizes $L=10$ and $20$, respectively, both for the
non-interacting (open circle) and interacting (solid circle) case.
The LDOS has been calculated at the fixed energy $E=-2t$, slightly
above the upper edge of the band (remember that at the parameter
values chosen below Eq.~(\ref{moms:19}), $E=-2t$ is the upper edge of
the band for an infinite tight-binding chain, i.e., $H_d$ with
$L=\infty$ and $U=0$ in Eq.~(\ref{moms:2})). The LDOS decays
exponentially with distance from the metal-molecule contact, except
for the region close to the boundaries. This reproduces the
exponential decay of the STM tunneling conductance in
Ref.~\cite{gimzewski99}. It also agrees with the expectation that the
electron wave function induced by the metal to the molecule should
decay in an exponential way. Another interesting conclusion that can
be drawn from Fig.~\ref{moms:fig4} is the influence of the interaction
on the suppression of the LDOS. The characteristic decay length
decreases with the interaction $U$. This can be interpreted by saying
that the strong repulsion by the electrons sitting on
(singly-occupied) sites tends to block tunneling events from the metal
to the molecule.  This effectively leads to the suppression of the
``hybridization'' term $H_T$, Eq.~(\ref{moms:3}). The suppression of
the LDOS by interaction effects does not depend on the choice $E=-2t$,
however, the spatial scale would be different for different positions
of the molecular level.  The influence of repulsive interactions was
also studied in \cite{favand98}, where it is shown that the
suppression is lower for repulsive interactions than for a
dimerization leading to the same value of the gap.

In conclusion, we have calculated the local density of states of a
molecule adsorbed to a metal surface using a new Anderson-type
model. The LDOS decays along the molecule in an exponential way (like
in the experiment \cite{gimzewski99}) and is suppressed by a local
interaction on the molecule. Our formalism provides a qualitative
understanding of this and similar experiments. It is able to include
local interaction effects and can be used to treat more complicated
geometries that will be investigated in future experiments.

\stars
We would like to acknowledge discussions with T.~Jung and A.~Baratoff 
and thank J.~Gimzewski for providing a preprint of Ref.~\cite{gimzewski99}.


\vskip-12pt
\begin{thebibliography}{99}

\bibitem{curacao} {\it Mesoscopic Electron Transport},
edited by L.~L. Sohn, L.~P. Kouwenhoven, and G. Sch\"on
(Kluwer, Dordrecht, 1997).

\bibitem{aviram74} A. Aviram and M.~A. Ratner, Chem. Phys. Lett. 
{\bf 29}, 277 (1974).

\bibitem{petty95} {\it Introduction to Molecular Electronics}, edited
by M.~C. Petty, M.~R. Bryce, and D. Bloor (Oxford University Press,
Oxford, 1995).

\bibitem{aviram98} {\it Molecular Electronics: Science and Technology},
edited by A. Aviram and M. Ratner, Ann. N.Y. Acad. Sci. {\bf 852} (1998).

\bibitem{eigler} D.~M. Eigler and E.~K. Schweizer, Nature {\bf 344}, 524 
(1990).

\bibitem{avouris} I.-W. Lyo and P. Avouris, Science {\bf 253}, 173 (1991).

\bibitem{crommie} M.~F. Crommie, C.~P. Lutz, and D.~M. Eigler, 
Science {\bf 262}, 218 (1993).

\bibitem{dekker1} S.~J. Tans {\it et al.},
Nature {\bf 386}, 474 (1997);
S.~J. Tans, R.~M. Verschueren, and C. Dekker,
Nature {\bf 393}, 49 (1998). 

\bibitem{bachtold} 
A. Bachtold {\it et al.}, Appl. Phys. Lett. {\bf 73}, 274 (1998).

\bibitem{gimzewski98} C. Joachim, J.~K. Gimzewski, and H. Tang,
Phys. Rev. B {\bf 58}, 16407 (1998).

\bibitem{ratner94} V. Mujica, M. Kemp, and M.~A. Ratner,
J. Chem. Phys. {\bf 101}, 6849 (1994); 
6856 (1994). 

\bibitem{joachim96} C. Joachim and J.~F. Vinuesa, 
Europhys. Lett. {\bf 33}, 635 (1996).

\bibitem{magoga9798} M. Magoga and C. Joachim, 
Phys. Rev. B {\bf 56}, 4722 (1997);
Phys. Rev. B {\bf 57}, 1820 (1998).

\bibitem{datta} S. Datta and W. Tian, 
Phys. Rev. B {\bf 55}, R1914 (1997).

\bibitem{kirczenow} E.~G. Emberly and G. Kirczenow, 
Phys. Rev. B {\bf 58}, 10911 (1998);
see also cond-mat/9908392.

\bibitem{ciraci} A. Buldum, D.~M. Leitner, and S. Ciraci,
Europhys. Lett. {\bf 47}, 208 (1999); A. Buldum, S. Ciraci, and
C.~Y. Fong, cond-mat/9908204. 

\bibitem{kergueris} C. Kergueris {\it et al.},
Phys. Rev. B {\bf 59}, 12505 (1999).

\bibitem{dattapaper} Y. Xue and S. Datta, cond-mat/9908073.

\bibitem{gimzewski99} V.~J. Langlais {\it et al.}, 
Phys. Rev. Lett. {\bf 83}, 2809 (1999).

\bibitem{gourdon98} A. Gourdon and H. Tang, Ann. N.Y. Acad. Sci.
{\bf 852}, 219 (1998).

\bibitem{Negele88}
J.~W. Negele and H. Orland, {\em Quantum Many-Particle Systems}
 (Addison-Wesley, Redwood, 1988);
E. Fradkin, {\em Field Theories of Condensed Matter Systems}
 (Addison-Wesley, New York, 1991).

\bibitem{dimerization} In practice, the band gap is due to both
correlation and dimerization effects. In this work, we concentration
on the correlation effects.

\bibitem{Liebxx68}
E.~H. Lieb and F.~Y. Wu, Phys. Rev. Lett. {\bf 20}, 1445 (1968).

\bibitem{Wiegma81}
P.~B. Wiegmann, Phys. Lett. A {\bf 31}, 163 (1981);
N. Kawakami and A. Okiji, Phys. Lett. A {\bf 86}, 483 (1981).

\bibitem{George96}
A. Georges, G. Kotliar, W. Krauth, and M.~J. Rozenberg,
 Rev. Mod. Phys. {\bf 68}, 13 (1996).

\bibitem{Auerba94}
A. Auerbach, {\em Interacting Electrons and Quantum Magnetism}
 (Springer-Verlag, Heidelberg, 1994).

\bibitem{Brinkm70}
W.~F. Brinkman and T.~M. Rice, Phys. Rev. B {\bf 2}, 1324 (1970).

\bibitem{favand98} J. Favand and F. Mila,
Eur. Phys. J. B {\bf 2}, 293 (1998).

\end{thebibliography}
\end{document}